\documentclass[
 superscriptaddress,
 preprint,
 reprint,
 bibnotes,
 showkeys,
 amsmath,amssymb,
 aps,
 prl,
]{revtex4-1}

\usepackage{graphicx}
\usepackage{amsmath}
\usepackage{dcolumn}
\usepackage{bm}
\usepackage{braket} 
\usepackage{bbold}
\renewcommand{\vec}{\bm}
\usepackage[
 colorlinks=true,
 urlcolor=blue,
 citecolor=blue,
 linkcolor=blue,
 bookmarks=false,
 pdfstartview={FitH},
]{hyperref}

\def\cm-1{cm$^{-1}$\,}
\def\cmT-1{cm$^{-1}$/T\,}
\def\B2{$B_{2g}$}
\def\B1{$B_{1g}$}
\def\Ts{$T_{S}$\,}

\begin{document}
\title{On the origin of the electronic anisotropy in 
 iron pnicitde superconductors } 
\author{W. -L.~Zhang}
\email{wlzhang@physics.rutgers.edu}
\affiliation{Department of Physics $\&$ Astronomy, Rutgers University, Piscataway, New Jersey 08854, USA}
\affiliation{Beijing National Laboratory for Condensed Matter Physics and Institute of Physics, Chinese Academy of Sciences, Beijing, 100190, China} 
\author{P.~Richard}
\affiliation{Beijing National Laboratory for Condensed Matter Physics and Institute of Physics, Chinese Academy of Sciences, Beijing, 100190, China}
\affiliation{Collaborative Innovation Center of Quantum Matter, Beijing, China}
\author{H.~Ding}
\affiliation{Beijing National Laboratory for Condensed Matter Physics and Institute of Physics, Chinese Academy of Sciences, Beijing, 100190, China}
\affiliation{Collaborative Innovation Center of Quantum Matter, Beijing, China}
\author{Athena~S.~Sefat}
\affiliation{Materials Science and Technology Division, Oak Ridge National Laboratory, Oak Ridge, Tennessee 37831-6114, USA}
\author{J.~Gillett}
\affiliation{Cavendish Laboratory, Cambridge University, JJ Thomson Avenue, Cambridge CB3 OHE, UK} 
\author{Suchitra~E.~Sebastian}
\affiliation{Cavendish Laboratory, Cambridge University, JJ Thomson Avenue, Cambridge CB3 OHE, UK} 
\author{M.~Khodas}
 \email{maxim.khodas@gmail.com}
\affiliation{Department of Physics and Astronomy, University of Iowa, Iowa City, Iowa 52242, USA}
\author{G.~Blumberg}
 \email{girsh@physics.rutgers.edu}
\affiliation{Department of Physics $\&$ Astronomy, Rutgers University, Piscataway, New Jersey 08854, USA}
\affiliation{National Institute of Chemical Physics and Biophysics, Akadeemia tee 23, 12618 Tallinn, Estonia}

\begin{abstract}
We use polarization-resolved Raman spectroscopy to 
study the anisotropy of the electronic characteristics of the iron-pnictide parent compounds $A$Fe$_{2}$As$_{2}$ ($A$~=~Eu, Sr) .
We demonstrate that above the structural phase transition at \Ts the 
dynamical anisotropic properties of the 122 compounds are governed by the emergence of 
xy-symmetry critical collective mode foretelling a condensation 
into a state with 
spontaneously broken four-fold symmetry at a temperature $T^{*}$. 
However, the mode's critical slowing down is intervened by a 
structural transition at \Ts, about 80~K above $T^{*}$, resulting in
an anisotropic density wave state. 
\end{abstract}
\date{\today}
\maketitle

The properties of solids are defined by the symmetry of the 
underlying lattice. 
For the iron pnictide superconductors, the electronic nematicity, a 
tendency of the electronic ground state 
to deform spontaneously in proximity to the tetragonal to 
orthorhombic structural transition, remains the focus of intense research 
activity~\cite{Fradkin_Annurev2010, Davis_NatPhys2014, Fernandes_NatPhys2014}. 
In the 122 family of iron pnictides, below the 
structural transition temperature $T_S$ that breaks the four-fold 
rotational symmetry, 
the crystal lattice constants $a$ and $b$ differ by less 
than one percent while the anisotropy of the electronic characteristics,  
such as the DC resistivity and the optical conductivity, can be as large as 30 
percent~\cite{Tegel_JPCM2008,Fisher_Science2010, Degiorgi_PRB2014}.   
More importantly, even above $T_S$, where the underlying lattice 
remains tetragonal, large electronic nematic susceptibility has been 
suggested from resistivity anisotropy measurements in a ``zero strain'' 
limit~\cite{Fisher_Science2012} and from the observation of a splitting 
of the electronic  bands derived from degenerate iron 
orbitals~\cite{YiM_PNAS2011,Shimojima_PhysRevB.89.045101}.  
However, the origin of nematicity remains controversial because of the 
strong interplay of multiple degrees of freedom.           

Theoretical proposals suggest that either the magnetic moments of 
the iron spins, which order in close proximity to \Ts into collinear 
stripes~\cite{Fernandes_NatPhys2014, Xu_PRB2008}, or the iron 
$d$-orbitals, which below \Ts no longer have the four-fold symmetry, 
drive the nematicity~\cite{WeiKu_PRL2009, 
kontani_SSC2012, onari_PRL2012}.      
Because the tetragonal to orthorhombic transition occurs in close 
proximity to the magnetic ordering transition (at $T_{SDW}$) and 
leads to superconductivity with an unconventional order parameter, 
in-depth studies of  
symmetry and critical dynamics of nematic fluctuations are essential to 
gain an insight into both magnetism and superconductivity in 
pnictides~\cite{fang_PRB2008,Hirschfeld_RRP2011, 
Fernandes_NatPhys2014,Fernandes_PhysRevLett2013,Millis_2013PRL.111.127001}.      

In this letter we use Raman spectroscopy to study the evolution of 
electronic anisotropy for representative 
parent compounds of the iron pnictide superconductors. 
We demonstrate that critical long temporal collective orbital fluctuations 
of xy-symmetry develop in a broad range of temperatures, 
foretelling the new ground state of the same symmetry at temperature $T^{*}$. 
However, condensation into the emergent state is intervened by a 
first order transition at \Ts, about 80~K above $T^{*}$, 
characterized by a broken four-fold symmetry, a doubling of the  
unit cell, an anisotropic density wave, and a collinear spin stripe order. 
We propose a microscopic interpretation of the observed anomalous 
evolution of the Raman susceptibility, 
elucidate reported anisotropy of electronic characteristics above and 
below \Ts, 
and explain the implications on superconductivity for doped pnictide 
materials. 
  
Polarization-resolved inelastic light scattering probes excitations 
of prescribed symmetries. 
The temperature evolution of the electronic Raman susceptibility 
reveals the dynamics of collective excitations and  
provides an unambiguous identification of their symmetry. 
Unlike most other symmetry sensitive probes requiring   
moderate to strong external perturbations, 
such as magnetic \cite{Matsuda_Nature2012} or strain fields 
\cite{Fisher_Science2012,LuDai_Science2014}, 
the photon field used in Raman scattering is weak.
Thus, Raman spectroscopy presents an ideal tool to study the dynamics 
and symmetry of nematic fluctuations without introducing external 
symmetry breaking perturbations.  

\begin{figure}[!t]
\begin{center}
 \includegraphics[width=\columnwidth]{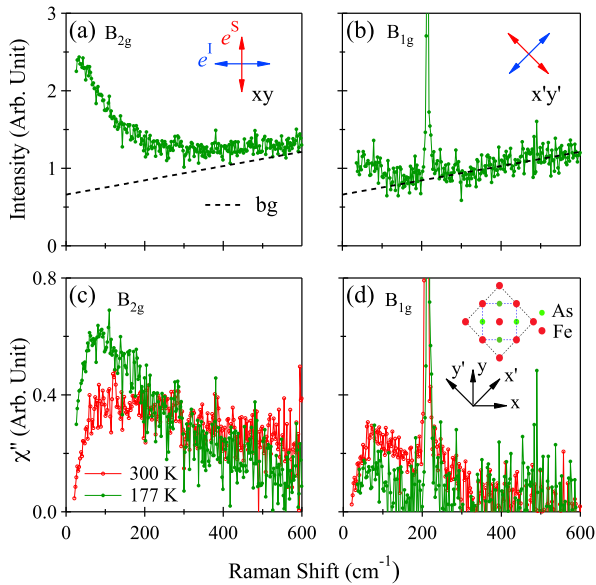}
\end{center}
\vspace{-3mm}
 \caption{\label{Fig: 1} (color online) 
 Secondary emission (a and b) and Raman response (c and d) for 
 EuFe$_2$As$_2$ at 177~K (green) and 300~K (red) for 
 xy (a and c) and 
 x$^{\prime}$y$^{\prime}$ (b and d) polarization geometries 
 corresponding to $B_{2g}$ and $B_{1g}$ responses in the 
 tetragonal phase, recorded with
 $\lambda_{L}$=476~nm. 
 Schematic diagram of the Fe-As layer is shown in (d), with the tetragonal 
 2-Fe unit cell in blue and the orthorhombic 4-Fe unit cell in black. 
 The light polarization vectors are denoted with 
 respect to the crystallographic orientations. 
 The luminescence background indicated by 
 black dotted lines in (a) and (b) is determined 
 in Refs.~\cite{supplementary,datafit}. 
 }             
\vspace{-3mm}
\end{figure}

The Raman response function  
traces electronic density-density correlations driven by 
the incident and scattered photons of the chosen polarizations  
($\vec{e}^I$ and $\vec{e}^S$ shown as blue and 
red arrows in Fig.~\ref{Fig: 1})
\begin{align}
    \label{corr_func}
\chi_{\vec{e}^I\vec{e}^S} (\omega) \propto - i  \int_0^{\infty} d t e^{ i \omega t}
\left\langle \left[\tilde{\rho}(t),\tilde{\rho}(0) \right] 
\right\rangle \, . 
\end{align}
In the effective mass approximation, $\tilde{\rho} = (m/\hbar^{2})
\sum_{\alpha,\beta}\mathrm{ e^I_{\alpha} e^S_{\beta}} \sum_{b,\vec{k}} 
\left[\partial^2 \epsilon^b_{\vec{k}}/\partial k_{\alpha} \partial k_{\beta} 
\right]  \hat{n}_{b\vec{k}}$, with $m$ denoting the electron mass 
and 
$\hat{n}_{b\vec{k}}$ the occupation of the Bloch state at momentum 
$\vec{k}$ in the band $b$ with the dispersion 
$\epsilon^b_{\vec{k}}$~\cite{Klein2010,Sauer_PRB2010,Mazin_PRB_2010}.     
All components of the symmetrized Raman tensor 
$\chi_{\mathrm{\vec{e}^I\vec{e}^S}}$ can be classified by irreducible 
representations of the crystallographic point group 
\cite{Ovander1960}. 
The two photon fields with cross polarization couple 
to quadrupole moments, and thus the scattering experiments probe the
$B_{2g}$ susceptibility if  
light polarization is aligned along $a$ and $b$ crystallographic 
directions, and the $B_{1g}$ susceptibility if rotated by 
$45^{\circ}$.  

\begin{figure*} [!t]
\begin{center}
\includegraphics[width=1.9\columnwidth]{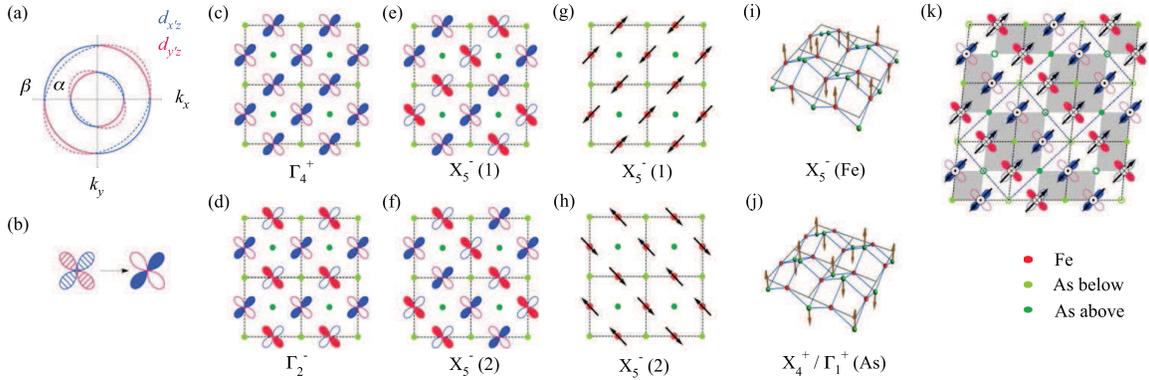}
\end{center}
\vspace{-3mm}
 \caption{\label{Fig: 3} 
(color online) 
(a) Schematic diagram of the $\alpha$ (inner) and $\beta$ (outer) 
Fermi surfaces at the BZ center $\Gamma$-point.  
The Fermi surfaces are distorted along the x$^\prime$ and y$^\prime$ 
directions due to the charge transfer fluctuations between the 
$d_{x^\prime z}$ and $d_{y^\prime z}$ orbitals. 
(b) A quadruple moment is caused by the charge transfer between the 
$d_{x^\prime z}$ and $d_{y^\prime z}$ orbitals. 
(c) and (d) Two possible long wavelength ($\Gamma$-point) 
configurations of quadrupoles with $\Gamma_4^+$($B_{2g}$) and $\Gamma_2^-$($A_{2u}$) symmetries. 
In the second Fe layer in the unit cell the $d_{x^\prime z}$ and 
$d_{y^\prime z}$ orbitals have the same/reversed occupation with the 
first layer for $B_{2g}$ and $A_{2u}$, respectively.  
(e) and (f) The degenerated quadruple excitations that belongs to the 
$X_{5}^{-}$-symmetry at the  corner.  
The second Fe layer has the reversed with the first layer orbital occupation. 
(g) and (h) The spin triplet excitation that belongs to the $X_{5}^{-}$-symmetry at the BZ corner. 
The second Fe layer has the reversed spin with the first layer.
(i) and (j) The atomic displacement of the  $X_{5}^{-}$ type of Fe 
and  $X_{4}^{+}$($D_{4h}$ phase)/$\Gamma_{1}^{+}$($D_{2h}$ phase) type of As shown in one Fe layer. 
The second Fe layer has the anti-phase/in-phase displacement with the first layer.
(k) The possible quadruple, spin and lattice configuration in the orthorhombic ground state below \Ts. 
The Fe $X_{5}^{-}$-symmetry displacement. 
Dots and crosses denote the Fe $X_{5}^{-}$-symmetry displacement up 
and down along the $c$-direction. 
The orthorhombic unit cell is shown by the blue dashed line. 
As ions above and below the Fe layer are distinguished in different colors. 
The shaded grey rhombuses show a checkerboard charge modulation of 
approximate xy symmetry. 
 }
\vspace{-0mm}
\end{figure*}

For the Fe pnictides, 
a local charge transfer between degenerate $d_{x^{\prime}z}$ and 
$d_{y^{\prime}z}$ orbitals without a spin flip induces a quadrupole 
moment of $B_{2g}$ symmetry on an iron site, Fig.~\ref{Fig: 3}(b), 
while no low energy excitation of $B_{1g}$ symmetry  
is allowed without charge transfer between the irons.  
Therefore, $\chi_{\mathrm{xy}}$, which bears four nodes  
along the crystallographic directions, delineates such singlet 
quadrupole excitations~\cite{Sauer_PRB2010,Mazin_PRB_2010}. 
An excitation with a spin flip excites a magnon 
to which, in the leading order, light cannot couple. 
In Figs.~\ref{Fig: 3}(a-d) we illustrate the relevant quadrupole excitation 
transition both in $k$-space and real space.

\begin{figure*} [!t]
\begin{center}
 \includegraphics[width=1.98\columnwidth]{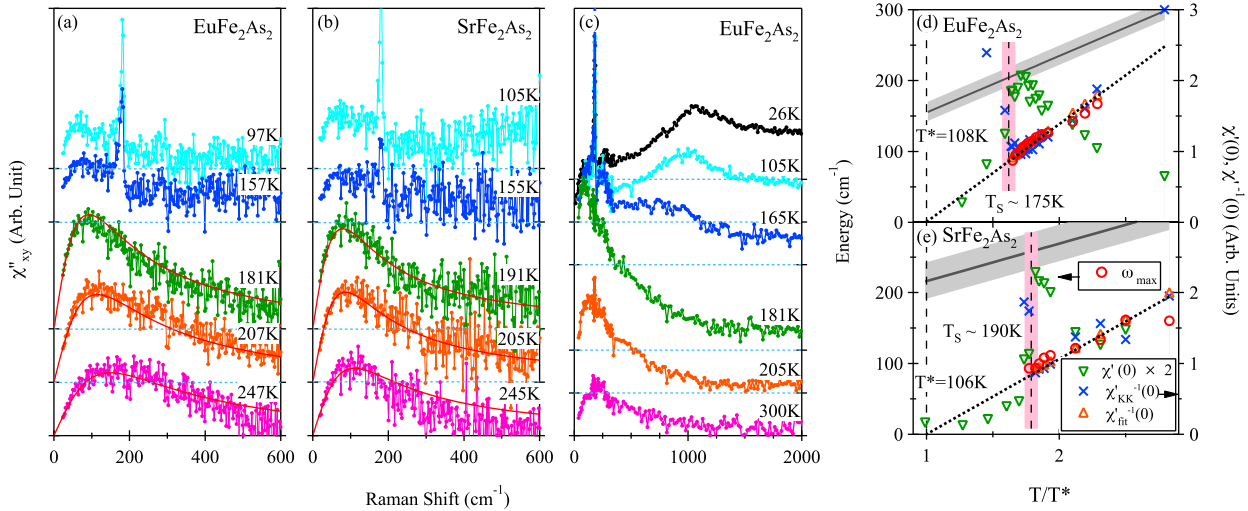}
\end{center}
\vspace{-3mm}
 \caption{\label{Fig: 2} (color online) 
 Evolution of the Raman response for the xy polarization with cooling 
 from the tetragonal into the orthorhombic phase 
(a and c) for EuFe$_2$As$_2$ and (b) for SrFe$_2$As$_2$. 
 The red curves are the fits to the phenomenological model  
 Eq.~\eqref{delta_XY}~\cite{datafit,supplementary}. 
 The data show the development of an anomalous quasi-elastic peak  
 above \Ts, its abrupt disappearance at \Ts, 
 followed by the formation of a density wave gap with a coherence mode at 
about 1070~\cm-1 and a new phonon at 183~\cm-1.  
 Laser excitations $\lambda_{L}$ of 476~nm (a and b) and 647~nm (c) were used. 
The temperature dependence of the 
$\chi_{\mathrm{xy}}^{\prime\prime}(\omega,T)$ response maximum 
$\omega_{max}(T)$ (red circles), of the  
relaxation rate $\Gamma_T$ (solid lines, shaded areas indicate the 
fitting error), the static Raman susceptibility 
$\chi_{\mathrm{xy}}^\prime(0,T)$ (green triangles), 
$\chi_{\mathrm{KK}}^{\prime-1}(0,T)$
 from data (blue crosses), and 
$\chi_{\mathrm{fit}}^{\prime-1}(0,T)$
from   
fit (orange triangles) for EuFe$_2$As$_2$ 
(d) and SrFe$_2$As$_2$ (e). 
The horizontal axis is scaled by $T^*$. 
The vertical black dashed lines at $T/T^*$\,=\,1.6 and 1.8 denote 
\Ts, while the red shaded areas indicate an hysteresis. 
 }
\vspace{-0mm}
\end{figure*}

In Figs.~\ref{Fig: 1}(a) and \ref{Fig: 1}(b) we compare the intensity measured in xy 
and x$^{\prime}$y$^{\prime}$ geometries~\cite{supplementary} 
from EuFe$_2$As$_2$ pnictide just above $T_S = 175$~K.
The Raman response shown in Figs.~\ref{Fig: 1}(c), \ref{Fig: 1}(d) and 
\ref{Fig: 2}
is derived from the scattering intensity by 
accounting for the Bose factor~\cite{datafit}. 

As shown in Fig.~\ref{Fig: 1}(d), 
the electronic response from intraband excitations using the 
x$^{\prime}$y$^{\prime}$ polarization is indeed very weak.   
The broad feature centered at about 75~\cm-1 is likely due to an 
interband transitions between the $\alpha$ and $\beta$ bands, as illustrated in   
Fig.~\ref{Fig: 3}\,(a). 
The sharp mode at 214~\cm-1 is an iron \B1 
phonon~\cite{Litvinchuk_PRB2008} which only shows weak anomaly upon 
cooling across \Ts \cite{supplementary}.

In contrast, the response for the xy polarization is much stronger, the 
signal extends beyond 1000~\cm-1, it strengthens significantly upon 
cooling and develops a well defined maximum at low energy.  
In Fig.~\ref{Fig: 2} we show that the temperature evolution of the 
$B_{2g}$ Raman susceptibility for EuFe$_2$As$_2$ and 
SrFe$_2$As$_2$ is generic for the 122 compounds~\cite{Gallais_PRL2013}.
Upon cooling toward \Ts, the data show gradual enhancement 
of the low energy response along with  linear in 
temperature softening of the mode maximum frequency from about 160~\cm-1 at room  
temperature to 90~\cm-1 at \Ts, denoted $\omega_{max}(T)$ in 
Figs.~\ref{Fig: 2} (d-e), an energy scale independent on the quasi-particle 
characteristics. 
No enhancement with cooling is observed for the susceptibility in any 
other symmetry channel. 
Such critical evolution of the anomalous $B_{2g}$ susceptibility due to 
quadrupole deformation of the electron density at the Fermi surfaces
is abruptly intervened at \Ts, the entrance into the 
orthorhombic phase, when a density wave like gap and a 
coherence peak at about 1070~\cm-1  develops.
The power law in the density wave gap structure suggests that the gap 
is anisotropic, possibly with a $d_{xy}$ symmetry. 

From these data, we calculate and 
plot in Figs.~\ref{Fig: 2} (d-e) the real part of the  
static Raman susceptibility using the Kramers-Kronig relations. 
${\chi_{\mathrm{xy}}^\prime(0,T)} \propto (T-T^*)^{-1}$ 
shows a mean-field-like enhancement 
with temperature $T^*$ about 80~K below \Ts.  
Remarkably, the mode's maximum frequency $\omega_{max}(T)$ scales to the very 
same $(T-T^*)$ behavior and does not soften near 
\Ts, indicating that 
the critical slow down is intervened abruptly at \Ts. 

The $B_{2g}$-susceptibility data cannot be understood in a 
picture of non-interacting incoherent quasi-particle excitations 
because in a paramagnetic phase such excitations can only produce 
a featureless low frequency response. 
The spectrum of low-frequency anisotropic plasmas is 
typically featureless and has a  
cutoff at $4\pi \hbar v_{F}/\lambda_{L}$ due to kinematic constraints 
of the scattering process, where $\lambda_{L}$ is the excitation 
wavelength and $v_{F}$ is Fermi velocity in the direction 
of photon propagation, relatively small for the layered iron 
pnictides~\cite{Platzman1965,Klein_PhysRevB1984}.
Hence, the presented 
${\chi_{\mathrm{xy}}^{\prime\prime}(\omega,T)}$  
data showing the emergence and softening of an anomalous 
quasi-elastic response above \Ts is a manifestation of electronic 
correlations.  

To describe the observed collective behavior we use an expression for 
interacting susceptibilities
\begin{align}\label{delta_XY}
\chi_{\mathrm{xy}}(\omega,T) \propto  
\frac{\chi^{(0)}_{\mathrm{xy}}(\omega,T)}{1-g 
\chi^{(0)}_{\mathrm{xy}}(\omega,T)}\,  
\end{align}
where $\chi^{(0)}_{\mathrm{xy}}$ is the non-interacting   
susceptibility, 
and $g$ is the coupling  
constant.
We conjecture 
\begin{align}\label{log}
\chi^{(0)}_{\mathrm{xy}}(\omega,T) = \frac{C}{\pi}\log\frac{ (\omega 
+ i \Gamma_T)^2 - \Lambda^2 }{\left(\omega + i \Gamma_T \right)^2 } 
\, ,   
\end{align}
as this form yields expected featureless bare response 
$\chi^{\prime\prime(0)}_{\mathrm{xy}}(\omega,T) \propto 
\arctan{(\omega/ \Gamma_{T})}$. 
Here $\Gamma_T$ describes a temperature dependent electron-hole 
relaxation rate $\gtrsim 2k_{B}T$ and $\Lambda$ is the 
ultraviolet cutoff.  
At $\omega \lesssim \Gamma_T$ Eqs.~\eqref{log} and \eqref{delta_XY} 
yield the relaxational form 
\begin{align}\label{q-Lor}
\chi^{\prime\prime}_{\mathrm{xy}}(\omega,T) \propto  
\omega \Gamma_T/ (\omega_{max}(T)^2 +  \omega^2)\,  ,     
\end{align}
with $\omega_{max}(T) = \Gamma_T [1/\tilde{g} -  \log ( 
\Lambda/\Gamma_T)]$ and $\tilde{g} = C g$. 
Because $\chi^{(0)}_{\mathrm{xy}}(0,T)$ is 
logarithmically large, 
Eq.~\eqref{delta_XY} guarantees the observed critical temperature 
dependence of the static susceptibility 
$1/\chi_{\mathrm{xy}}(0,T) \propto 
(1/\chi^{(0)}_{\mathrm{xy}}(0,T) -g)$ with $T^*$ defined by the 
coupling constant $g$,  
$\chi^{(0)}_{\mathrm{xy}}(0,T^*) = 1/g$. 

The expression \eqref{delta_XY} enables us to fit the 
entire temperature 
evolution of the $\chi_{\mathrm{xy}}(\omega,T)$ susceptibility 
between the room temperature and \Ts \cite{supplementary}. 
The data show critical enhancement of the susceptibility and slow 
down of the characteristic fluctuation frequency $\omega_{max}(T)$, which is 
abruptly intervened by the structural transition at $T_{S} \approx 1.6 
- 1.8\,\,T^*$ (See Fig.~\ref{Fig: 2}). 
Above $T_{S}$, the observed anomalous temperature dependence of the $B_{2g}$ susceptibility 
is arising from the critical collective fluctuations indicating the 
system's approach to a phase transition breaking the four-fold symmetry.  
The susceptibility reveals how, counteracting the relaxation 
processes, long temporal correlations characterized by the same 
symmetry develop, foretelling inception of the new ground state at 
temperature $T^{*}$.   

We now turn to the microscopic interpretation of the observed anomalous 
$\chi_{\mathrm{xy}}(\omega,T)$. 
The body centered unit cell of 122 structure contains two Fe layers 
with two Fe sites in each one. 
In the long wavelength limit, $B_{2g}$ and $A_{2u}$ 
are two possible symmetries of the quadruple excitations, as illustrated in Fig.~\ref{Fig: 3}(c-d).
The only Raman 
active quadrupole excitations are singlets of $\Gamma_{4}^{+}$
($B_{2g}$) symmetry.  
Those quadrupole excitations precipitate the observed anomalous 
susceptibility. 
The critical behavior of $\chi_{\mathrm{xy}}(\omega,T)$ foretells that the system 
contemplates a phase transition at temperature $T^{*}$ which would break 
the four-fold symmetry without a structural density wave instability 
and establish the ground state depicted in Fig.~\ref{Fig: 3}(c).

For the momentum at the Brillouin zone (BZ) corner, 
the \Ts structural 
instability vector 
$X=(\pi/a, \pi/a, 2\pi/c)$, both  
the singlet quadrupole excitations and the spin triplet 
excitations obey a   
degenerate $X_{5}^{-}$ symmetry, as illustrated in Figs.~\ref{Fig: 3}(e-h). 
The quadrupole-quadrupole and/or spin-spin interaction between nearest and 
next-nearest Fe sites 
favors such collinear stripe order as its ground 
state~\cite{chandra_PRL1990,Fernandes_PRL2010}. 
Coupling of quadrupole and spin excitations to the lattice leading to a structural 
phase transition is naturally expected for non-symmetric fluctuations. 
If, as it is here, the electronic state is degenerate and can interact 
with a degenerate lattice displacement, the Jahn-Teller theorem 
predicts the lifting of the $X_{5}^{-}$ quadrupole state degeneracy by interaction with 
$X_{5}^{-}$ phonons~\cite{Bersuker_2001}. 
The Fe $X_{5}^{-}$ phonon mode, illustrated in Fig.~\ref{Fig: 3}(i), buckles the flat 
Fe layer, modulates the distance between quadrupoles  
and/or spins and therefore promotes the choice of quadrupole and spin collinear stripe order.  
The solution of vibronic cooperative $X_{5}^{-} \otimes X_{5}^{-}$ Jahn-Teller 
problem results in a new ground state shown in Fig.~\ref{Fig: 3}(k). 
When the lattice symmetry breaks, the quadrupole and spin collinear 
stripe orders lock-in. 
Hence, for the parent compounds, the first order structural transition with 
doubling of the unit cell at \Ts, 80~K above $T^{*}$, intervenes the 
formation of the contemplated long-range ordered electronic state.  
The new order induces long-range charge modulation observed at low 
temperatures as the $d_{xy}$-symmetry density wave gap. 
The distortion of the As layer is induced by coupling to the new electronic 
ground state of the same symmetry, a quadrupole- and/or 
magneto-striction, and is revealed by the appearance of the new Raman 
active As phonon mode below \Ts in the spectra of $xy$ polarization, 
Fig.~\ref{Fig: 2}(a-c) and \ref{Fig: 3}(j). 
As the lattice modulations are buckle-like, the induced anisotropy of 
the lattice parameters remains weak.  

The results obtained here for parent compounds might bear implications on 
superconductivity in the Fe pnictides. 
As  superconductivity evolves from the tetragonal phase where the 
$B_{2g}$-type fluctuations remain dynamic and strong, they 
may mediate the superconducting pairing of unconventional symmetry, 
or appear as long-lived in-gap collective modes. 

In summary, we use polarization-resolved Raman spectroscopy to study 
the emergence and the competition of the collective orbital, spin and 
lattice fluctuations in EuFe$_2$As$_2$ and 
SrFe$_2$As$_2$ parent compounds of the iron pnictide 
superconductors. 
In the xy scattering geometry the Raman 
response exhibits a broad excitation spectrum with a peak at low 
energy showing critical behavior in a wide temperature range above 
the structural transition.     
This spectrum reveals dynamics of the relaxational mode due to 
orbital fluctuations of $B_{2g}$-symmetry and demonstrates the general 
interrelation between anticipated phase transitions breaking the discrete 
symmetry and critical fluctuations characterized by the 
same symmetry.  
For the studded parent compound the condensation of the 
emergent $B_{2g}$-mode is intervened by a  
structural density wave transition characterized by a doubling of the 
unit cell and a collinear spin stripe order.

We thank Y.~Gallais, A.~Sacuto, V.-K.~Thorsmolle, and Z.P.~Yin for 
fruitful discussions. 
W.-L.Z. acknowledges ICAM (NSF-IMI grant DMR-0844115) and NSF (DMR-1104884).
M.K. acknowledges support by the University of Iowa. 
P.R. and H.D. acknowledge MoST (2011CBA001001) and NFSC (11274362) of China. 
S.E.S. acknowledges ERC (FP/2007-2013)/ERC Grant Agreement no. 337425.
A.S.S. and G.B. acknowledge the US DOE, BES and Division of Materials 
Sciences and Engineering under Awards to ORNL and {DE-SC0005463} correspondingly.


\begin{thebibliography}{32}
\bibitem [{dat()}]{datafit}%
  \BibitemOpen
  \href@noop {} {}\bibinfo {note}
  {$I_{\mathrm{xy}}(\omega,T)=\chi_{\mathrm{xy}}^{\prime\prime}(\omega,\,T)(1+n_{\omega,T})+
  I_{bg}$ is used to describe the Raman response and the background for spectra
  at all temperatures. $n_{\omega,T}$ is the Bose Einstein distribution
  temperature factor. The background function $I_{bg}=a+b \omega+ cT$ while
  $a$, $b$ and $c$ are fitting parameters. For one particular compound all the
  coefficients namely $a$, $b$, $c$, $\alpha$ and $\Gamma_{\mathrm{res}}$ are
  kept constant for all temperatures in the nematic fluctuating
  phase.}\BibitemShut {Stop}%
\end{thebibliography}
\end{document}